Full Length Paper

**Title:**

Electrical Properties of RF-sputtered $0.65Pb(Mg_{1/3}Nb_{2/3})O_3$–$0.35PbTiO_3$ Thin Films on $La_{0.67}Ca_{0.33}MnO_3$ Buffered Platinized Glass Substrate

Authors:

*T.Garg[1,2], A. R. Kulkarni[1] and N. Venkataramani[1]

 **Affiliations:**

[1]Department of Metallurgical Engineering & Materials Science, Indian Institute of Technology Bombay, India-400076

[2]Department of Physics, College of Engineering Studies, University of Petroleum & Energy Studies, Dehradun, India-248007

Corresponding author:

Tarun Garg

Department of Metallurgical Engineering & Materials Science

Indian Institute of Technology Bombay

Powai, Mumbai, India-400076

Email: gargphy@iitb.ac.in

Contact No.: +91-22-25767657

# Research Highlights

- Single phase 0.65PMN-0.35PT thin films were deposited by RF- magnetron sputtering
- $La_{0.67}Ca_{0.33}MnO_3$ buffered platinized glass was used as substrate
- Effect of annealing temperature on structural and electrical properties is studied
- Comparable values for electrical properties could be achieved on buffered glass

# Abstract

Lead based relaxor ferroelectric thin films have been of technological importance due to excellent properties for several commercial applications. However, subtleness and high cost of fabrication have plagued of these materials. In this work, a systematic study of $0.65Pb(Mg_{1/3}Nb_{2/3})O_3$–$0.35PbTiO_3$ (PMN-PT) thin films grown on $La_{0.67}Ca_{0.33}MnO_3$ (LCMO) buffered Pt/$TiO_2$/Glass substrates has been undertaken. The films were grown at room temperature using RF- magnetron sputtering. Single phase PMN-PT films could be obtained by *ex-situ* thermal annealing in air for 2h at temperatures 550 and 650 °C. The films annealed at temperatures lower than 550°C and films without buffer layer showed presence of pyrochlore. Effect of annealing temperature on the microstructure, dielectric and ferroelectric properties of the PMN-PT films has been investigated. Scanning electron micrographs of single phase PMN-PT films show a bimodal grain size distribution for films annealed at 550 and 650 °C. A high dielectric constant of 1300 and a remnant polarization ($2P_r$) of 17μC/$cm^2$ were observed for film annealed at 650 °C.

# 1. Introduction

Bulk relaxor ferroelectric materials $(1-x)Pb(Mg_{1/3}Nb_{2/3})O_3-xPbTiO_3$ (PMN-PT) with compositions around the morphotropic phase boundary($0.33<x<0.37$) show excellent dielectric and piezoelectric properties[1-2]. Thin films of these materials are being explored for microelectromechanical systems (MEMS) and ferroelectric nonvolatile random access memory (FeRAM) applications[3-4]. Fabrication of these films poses a challenge of reproducibility of single phase films with good dielectric and ferroelectric properties. This challenge has made these materials topic of great research interest. One major issue with this system is the formation of the undesired pyrochlore phase during the synthesis both in bulk and thin film form. Even a small amount of pyrochlore phase deteriorates the dielectric and ferroelectric properties of PMN-PT drastically [5]. In bulk the problem was resolved using a two-step columbite method [6]. For the thin films this method is difficult to implement. Other methods suggested for suppressing the pyrochlore phase in thin films are control of the lead(Pb) content [7-8], introduction of a buffer or seed layer [9-17] or using a single-crystal substrate [18-20].

Lead content can be controlled by using targets with Pb excess such as in sputtering, pulsed laser deposition or using solutions with Pb ion excess in sol-gel technique. However, there is an ambiguity regarding how much excess Pb should be incorporated for pure perovskite phase formation and has not always worked.

A buffer or seed layer could be quite useful for preventing formation of pyrochlore. Additionally it can be effective in improving electrical properties of PMN-PT layer due to good lattice matching. It also works as a diffusion barrier between the ferroelectric thin film and the bottom electrode [9]. Complete elimination of pyrochlore phase is found quite difficult without buffer layer on metal coated substrates [7,21]. High activation energy for nucleation of perovskite phase has been found to be main reason for the pyrochlore phase formation which can be reduced by using a seed or buffer layer. A buffer layer reduces this nucleation barrier by providing heterogeneous sites for nucleation of perovskite phase [12]. Several reports on various perovskite and non- perovskite buffer layers such as PZT, PbTiO$_3$, BaTiO$_3$, La$_{1-x}$Sr$_x$CoO$_3$, La$_{1-x}$Sr$_x$MnO$_3$, TiO$_2$ and PbO used for PMN-PT films are available in literature where introduction of a buffer layer has not only suppressed the pyrochlore formation but also improved the structural and electrical properties [9-16]. Influence of PbTiO$_3$(PT), BaTiO$_3$(BT) and Ba$_{0.8}$Sr$_{0.2}$TiO$_3$(BST) buffer layers on structural and dielectric properties of PMN-PT films was studied by Nakamura *et. al.* BST buffered films showed 98% perovskite phase formation and a dielectric constant of 1270 [9]. Arai *et. al.* reported single perovskite phase formation in PbTiO$_3$ buffered PMN-PT films on Pt/TiO$_2$ coated silicon substrate along with a large remnant polarization(2P$_r$) of 62 μC/cm$^2$. However, 3% pyrochlore phase was observed on Pt/TiO$_2$ coated silicon without buffer layer[10]. An enhancement in perovskite phase content along with improved microstructure and electrical properties was observed with introduction of a PZT interfacial layer between PMN-PT and bottom electrode by Yang *et. al*. [12]. Therefore, use of a buffer layer has been observed to be essential for improvement in structural and electrical properties of PMN-PT films.

The other suggested way of getting single phase PMN-PT films is use of single crystal substrates such as MgO, LaAlO$_3$ or SrTiO$_3$ that too with a buffer layer. Use of single crystal substrates or even silicon (Si) in comparison with glass or quartz, make these thin films expensive. There have been a few studies undertaken in past on PMN-PT films deposited on glass however not much success could be achieved. Kumar *et. al* reported a dielectric constant of

950 for 0.8 μm thick 0.68PMN-0.32PT films having pyrochlore deposited on ITO coated glass derived by sol-gel[22]. Jaydeep *et. al* reported a dielectric constant of 2052 for 800 nm thick PMN films deposited on gold (Au) coated glass substrates by RF-magnetron sputtering[23].

Keeping all these factors in mind such as effect of buffer layer and high cost of single crystal substrate, we have attempted deposition of PMN-PT films by radio frequency magnetron sputtering on LCMO buffered 7059 corning glass substrate having Glass/ $TiO_2$/Pt configuration. LCMO has a perovskite crystal structure and therefore, well serves as a buffer layer for nucleation of perovskite phase of PMN-PT films. This study was aimed to fabricate single phase PMN-PT films on glass substrate with comparable electrical properties. The advantages offered by a buffer layer motivated us to use it on inexpensive glass substrates. To the best of our knowledge such a process has been shown for the first time. A systematic investigation on structural, dielectric and ferroelectric properties of PMN-PT films annealed at different temperatures have been carried out. Use of 7059 corning glass as substrate has also capped the temperature of annealing to 650 °C.

## 2. Experimental Details

0.65PMN-0.35PT (PMN-PT) stoichiometric target with 95% of the theoretical density was prepared using PbO, $Nb_2O_5$, $TiO_2$ and $4MgCO_3$. $Mg(OH)_2$. $4H_2O$ precursors by the columbite process to obtain a single phase perovskite. LCMO target was prepared using $La_2O_3$, $CaCO_3$ and $MnO_2$ precursors by conventional solid state reaction method with 94% of the theoretical density. Thin films were grown at room temperature using a three target RF magnetron sputtering (Edwards E306A) having a provision of substrate holder rotation. The substrate was platinum-coated 7059 corning glass having a configuration Glass/$TiO_2$(20nm)/Pt(100nm). LCMO and PMN-PT targets were mounted inside the vacuum chamber. A 3cm×2cm size substrate was placed at a distance of 6 cm above the target. The chamber was evacuated to a base pressure of $4.2\times10^{-6}$ mbar with a turbo molecular pump. First, an LCMO template layer of ~60 nm thickness was grown in $O_2$ plasma at a deposition pressure of $6\times10^{-3}$ mbar. Subsequently the substrate was moved over the PMN-PT target to deposit a 500 nm thick layer of PMN-PT. A deposition pressure of $2\times10^{-2}$ mbar was maintained using Ar/$O_2$ mixture in ratio 9:1 with the help of an MKS multigas mass flow controller (Model 647C) for PMN-PT layer. For all the thin film growth experiments 100 W of RF power is employed. As-prepared sample was cut in to 6 pieces of 1cm×1cm each for annealing at different temperatures and various characterizations. The as-deposited films were found X-ray amorphous. Therefore, post-deposition annealing was required to crystallize the films in the perovskite phase. The films were annealed in air at 450, 550 and 650°C for 2h and thereafter subjected to structural and electrical characterizations. In order to measure the ferroelectric and dielectric properties of the PMN-PT films, aluminium (Al) electrodes with the diameter of 0.25 mm and thickness of 100 nm were deposited by thermal evaporation using a shadow mask. A stacked capacitor configuration of Glass/$TiO_2$/Pt/LCMO/PMNPT/Al was achieved.

Phase and crystal structure of the as-sputtered and the annealed films were investigated by Philips PANalytical X-ray diffractometer using Cu-$K_\alpha$ ($\lambda$=1.54Å) radiation. The thicknesses of the various layers were measured using a surface profilometer (Bruker DektakXT). Thickness of PMN-PT film was around 500nm whereas thickness of LCMO layer was around 60 nm. The surface and cross-sectional morphologies of the films were examined by Field Emission Scanning Electron Microscopy (FE-SEM) (JEOL JSM-7600F). The dielectric and ferroelectric properties were studied using Novocontrol dielectric broadband spectrometer and *AIXacct* Ferroelectric Tester (TF2000 analyzer) respectively.

## 3. Results and Discussion

Figure 1 shows XRD patterns of as- deposited and annealed PMN-PT films recorded at room temperature. For comparison, a representative XRD pattern of the PMN-PT target is also been included in Figure 1. It is well known that low-temperature deposition is conducive to the formation of very fine-grained or even amorphous structures. As we have deposited films at room temperature, as-deposited PMN-PT films were found to be X-ray amorphous and only characteristic peaks of underlayers Pt and $TiO_2$ were observed in the XRD pattern. The films annealed at 550 and 650 °C showed single perovskite phase and are polycrystalline in nature. However, films annealed at 450 °C and below showed presence of pyrochlore phase which is favoured thermodynamically at temperatures below 500 °C. Increase in annealing temperature has pronounced effect on peak intensities of PMN-PT. The characteristic diffraction peaks of PMN-PT have become sharper and stronger in 650 °C annealed film when compared with 550 °C annealed film indicating grain growth with increase in annealing temperature. All the observed PMN-PT peaks could be indexed to perovskite phase using JCPDS card number 01-088-1853 while LCMO and $TiO_2$ peaks were indexed using JCPDS card numbers 00-049-0416 and 01-076-0323 respectively. The pyrochlore phase was identified as $Pb_2Nb_2O_7$ with orthorhombic crystal structure and the peak observed at 2θ≈29.5° could be indexed as (222) using JCPDS card number 00-040-828.

Figure 2 shows the cross-sectional SEM image of PMN-PT films. From cross-sectional view it can be seen that the PMN-PT, LCMO and Pt/Ti films have thicknesses around 500 nm, 60 nm and 120 nm respectively which are in agreement with the values measured using surface profilometer. Figure 3 shows the surface morphologies of the PMN-PT films annealed at 450, 550 and 650 °C. The effect of increase in annealing temperature is quite obvious from the microstructure of films shown in figure 3. It can be observed from figure 3(a) that in film annealed at 450 °C, crystallization has not been yet started due to unavailability of enough energy required for crystallization at that temperature. The observation of very fine grain structure corresponds to presence of pyrochlore phase [12]. This fact is also confirmed from X-ray diffraction pattern of the film. Further, as the annealing temperature is raised to 550 °C, perovskite phase crystallization occurred along with a heterogeneous grain growth. Some of the regions of the film are densely packed while some have lot of porosity as can be seen from Figure 3(b). This kind of abnormal grain growth in thin films is often manifested by a bimodal grain size distribution due to secondary grain growth in which some grains may grow very large (e.g., by a factor of ~ $10^2$) relative to surrounding ones [24]. Further increase in annealing temperature to 650 °C has caused increase in grain size resulting in improved density of the film. This is also reflected in X-ray diffraction pattern of 650 °C annealed film where characteristic peaks of PMN-PT have become more intense. Moreover, a bimodal grain size distribution can be observed in the film. The grain sizes calculated using average intercept length method[25] from scanning electron micrographs varied from 30-100 nm in 650 °C annealed film while in 550 °C annealed film, grain size varied from 10-40 nm.

Figure 4 shows the frequency dependence of the dielectric constant ($\varepsilon_r$) and loss tangent measured at room temperature in frequency range of 1Hz-1MHz. An excitation voltage ($V_{rms}$) of 0.1V was applied across the two electrodes of the film to perform the measurement. The dielectric constant of the films decreases with increase in frequency. This is due to space charge polarization, which is inherently related to the non-uniform charge accumulation. It can be observed from figure 4 that annealing temperature has significant effect on dielectric properties. The film annealed at 650 °C showed a dielectric constant of 1300 and dissipation factor of 0.02 whereas 550 °C annealed film showed a dielectric constant of 1000 and dissipation factor ~0.02 at 1 kHz. These values are comparable to those reported in literature where PMN-PT films were grown on buffered single crystal substrates such as MgO, $SrTiO_3$, $LaNiO_3$ and silicon [7, 18-21]. As for comparison the dielectric constant and saturation polarization values of PMN-PT thin film deposited on glass, $Pt/SiO_2/Si$ and buffered $Pt/SiO_2/Si$ substrates by different deposition techniques together with our experimental results are given in table 1. It can be seen from the table that our dielectric constant values are better than the reported values on glass and silicon substrates for the films deposited by RF sputtering. Further, for the films annealed below 550 °C, dielectric constant reduced drastically. For the film annealed at 450 °C, dielectric constant was ~120. As this film has pyrochlore as the dominant phase which has a low dielectric constant of ~130 in bulk form as compared to PMN-PT ceramics, the observed dielectric constant value is in fair agreement with the dielectric constant of pyrochlore phase [5]. For further decrease in annealing temperature to 350 °C, the dielectric constant reduced further. The decrease in dielectric constant value with decrease in annealing temperature can be attributed to presence of pyrochlore phase, smaller grain size and poor densification of the films which are also evident from X-ray diffraction patterns and microstructure of the films. It has been observed that the dielectric properties of the bulk crystal are significantly degraded in the thin films. Various factors such as deposition technique, processing temperatures, grain size, substrate, strain developed during the film growth and the formation of a thin low permittivity dielectric layer(called as dead layer) at the film electrode interface, may affect the dielectric properties of ferroelectric thin films substantially. The formation of dead layer causes high concentration of defects which cause domain wall pinning and also serves as trap centers for mobile charges [26].

Figure 5 shows P-E hysteresis loops of 550 and 650 °C annealed films recorded at different frequencies. Both these films showing single perovskite phase exhibited good ferroelectric hysteresis loops. However, for films annealed below 550 °C, the hysteresis loops were not well developed due to presence of large amount of pyrochlore phase. The effect of annealing temperature can be seen by comparing the P-E loops of two films. The film annealed at 650 °C showed a remnant polarization ($2P_r$) of 17 µC/cm$^2$ and a coercivity ($E_c$) of 115 V/cm while the film annealed at 550 °C showed a remnant polarization ($2P_r$) of 12 µC/cm$^2$ and a coercivity($E_c$) of 145 V/cm at 1kHz. Lower remnant polarization of 550 °C annealed film as compared to 650 °C annealed film could be attributed to smaller grain size. The ferroelectric properties are dependent on the ferroelectric domain structure, domain nucleation, and domain mobility. A correlation of the grain size with domain structure and the domain wall mobility of PbTiO$_3$ thin films based on the TEM observations was made by Ren *et al.*[31]. The grains having dimensions close to those of single domain have been found to be less influenced under an external field causing difficulty in domain nucleation. Consequently, the films with small grain size are single-domain predominated and show poor ferroelectric properties. Therefore, reduction in remnant polarization becomes obvious in ferroelectric films with small grains. On the other hand, the films with large grain sizes have a multi-domain predominated structure. This leads to easy switching of dipoles causing large remnant polarization. Thus, the ferroelectric properties are strongly dependent on the grain size. Moreover, the reduction in polarization can also be attributed to the pinning effect of the domain walls[32,33]. The smaller the grain size, more the number of grain boundaries, causing more pinning of domain walls hence reduces the domain wall mobility of ferroelectric thin films.

Further, large coercivity in our films may be attributed to large number of oxygen vacancies which can move into oxygen sublattice and form re-orientable dipoles with the impurity cations and defects. These dipoles stabilize the domain configuration by orienting themselves in direction of polarization vector which creates internal fields. These internal fields also cause the slight horizontal shift in P-E loops [11]. The asymmetry of the P-E loops of both the films is due to the difference in work functions of bottom and top electrodes which in our case are platinum and aluminum respectively. Since Pt and Al have different work functions which may cause a change in value of capacitance of PMN-PT capacitors under positive and negative bias.

Effect of increase in frequency of the bias field can be observed from decrease in saturation polarization ($P_s$) for both the films. At higher frequency, electric field changes its direction too fast for the electric dipoles to respond which causes decrease in saturation polarization at higher frequency. A high saturation polarization of $22\mu C/cm^2$ was observed for at 0.1 kHz which reduced to $18\mu C/cm^2$ at 1 kHz. A similar behavior was observed for 550 °C annealed film. A comparison of obtained $P_s$ values with the reported $P_s$ values for PMN-PT films deposited on glass substrate is given in the table 1.

## 4. Conclusions

For the first time, single phase PMN-PT films were prepared on LCMO buffered metal coated glass substrates using RF magnetron sputtering. LCMO buffer layer could prevent the formation of pyrochlore phase by stabilizing perovskite phase. Effect of annealing temperature on structural, dielectric and ferroelectric properties of PMN-PT films has investigated. A high dielectric constant of 1300 and high remnant polarization (2P$_r$) of 17 μC/cm$^2$ was observed for a film annealed at 650 °C. Use of buffer layer such as LCMO on metal coated glass substrates can lead to the preparation of single phase PMN-PT films with comparable electrical properties and hence may replace single crystal substrates and thus makes these films cost effective.

# References

1. G. Trefalt, B. Malic, J. Holc, H. Ursic, and M. Kose, "Synthesis of $0.65Pb(Mg_{1/3}Nb_{2/3})O_3$–$0.35PbTiO_3$ by Controlled Agglomeration of Precursor Particles" *J. Am. Ceram. Soc.* **95** (2012) 1858-1865.

2. M. Alguero, C. Alemanya, B. Jimeneza, J. Holcb, M. Kosecb and L. Pardoa, "Piezoelectric PMN-PT ceramics from mechanochemically activated precursors", *J. Euro. Ceram. Soc.* **24** (2004) 937–940.

3. D. L. Polla and L. F. Francis, "Ferroelectric Thin Films in Microelectromechanical System Applications," *Mat. Res. Soc. Bull.*, **21** (1996) 59–65.

4. J. F. Scott and C. A. Paz de Araujo, "Ferroelectric Memories," *Science*, **246** (1989) 1400–1405.

5. J. Chen and M. P. Harmer, "Microstructure and Dielectric Properties of Lead Magnesium Niobate-Pyrochlore Diphasic Mixtures", *J. Am. Ceram. Soc.* **73** (1990) 68-73.

6. Swartz, S.L. and Shrout, T., "Fabrication of Perovskite Lead Magnesium Niobate", *Mater. Res. Bull.*, **17**(1982)1245–1250.

7. M. C. Jiang and T. B. Wu, "The effect of electrode composition on rf magnetron sputtering deposition of $Pb[(Mg_{1/3}Nb_{2/3})_{0.7}Ti_{0.3}]O_3$ films", *J. Mater. Res.*, **9** (1994) 1879-1886.

8. H. Q. Fan, G. T. Park, J. J. Choi, H. E. Kim, "Preparation and Characterization of Sol—Gel-Derived Lead Magnesium Niobium Titanate Thin Films with Pure Perovskite Phase and Lead Oxide Cover Coat", *J. Am. Ceram. Soc.*, **85** (2002) 2001–2004.

9. T. Nakamura, A. Masuda, A. Morimoto and T. Shimizu, "Influence of Buffer Layers on Lead Magnesium Niobate Titanate Thin Films Prepared By Pulsed Laser Ablation", *Jpn. J. Appl. Phys.*, **35** (1996) 4750-4754.

10. T. Arai, Y. Goto, H. Yanagida, N. Sakamoto, T. Ohno, T. Matsuda, N. Wakiya and H. Sujuki, " Effects of Oxide Seeding Layers on Electrical Properties of Chemical Solution Deposition-Derived $Pb(Mg_{1/3}Nb_{2/3})O_3$–$PbTiO_3$ Relaxor Thin Films, *Jpn. J. Appl. Phys.*, **52** (2013) 09KA07-1–09KA07-5.

11. W. Gong, J-F. Li, X. Chu and L. Li, " Texture Control of Sol-Gel Derived Pb(Mg$_{1/3}$Nb$_{2/3}$)O$_3$–PbTiO$_3$ Thin Films Using Seeding Layer" *J. Am. Ceram. Soc.* **87** (2004) 1031-1034.

12. S. A. Yang, S. Y. Cho, J. S. Lim and S. D. Bu, "Distribution of pyrochlore phase in Pb(Mg$_{1/3}$Nb$_{2/3}$)O$_3$–PbTiO$_3$ films an suppression with a Pb(Zr$_{0.52}$Ti$_{0.48}$)O$_3$ interfacial layer", *Thin Solid Films*, **520** (2012) 7071-7075.

13. Q. Yao, F. Wang, C. M. Leung, Y. Tang, T. Wang and C. Jin, "Ferroelectric and dielectric properties of La$_{0.6}$Sr$_{0.4}$CoO$_3$-buffered 0.7Pb(Mg$_{1/3}$Nb$_{2/3}$)O$_3$–0.3PbTiO$_3$ thin film pulsed laser deposition", *J. Alloys Comp.*, **588** (2014) 290-293.

14. Z. Feng, D. Shi, R. Zeng and S. Dou, "Large electrocaloric effect of highly (100)-oriented 0.68Pb(Mg$_{1/3}$Nb$_{2/3}$)O$_3$–0.32PbTiO$_3$ thin films with a Pb(Zr$_{0.52}$Ti$_{0.48}$)O$_3$/PbO$_x$ buffer layer, *Thin Solid Films*, **519** (2011) 5433-5436.

15. N. Wakiya, K. Shinozaki and N. Mizutani, "Stabilization of Perovskite Pb(Mg$_{1/3}$Nb$_{2/3}$)O$_3$ thin film by a thin BaTiO$_3$ buffer layer on Pt/Ti/SiO$_2$/Si", *Thin Solid Films*, **409** (2002) 248-253.

16. J. Jiang and S-G. Yoon, "Epitaxial 0.65Pb(Mg$_{1/3}$Nb$_{2/3}$)O$_3$–0.35PbTiO$_3$ (PMN-PT) thin films grown on LaNiO$_3$/CeO$_2$/YSZ buffered Si substrates", *J. Alloys Comp.*, **509** (2011) 3065-3069.

17. J. M. Wang, W. L. Li, C. Q. Liu, W.D. Fei, "Effect of titanaium dioxide(TiO$_2$) seed layer on the phase composition of 0.7Pb(Mg$_{1/3}$Nb$_{2/3}$)O$_3$–0.3PbTiO$_3$ film prepared by pulsed laser deposition" *Microelectronic Eng.*, **85** (2008) 1920-1923.

18. V. Nagarajan, S. P. Alpay, C. S. Gunpule, B. K. Nagaraj, S. Aggarwal, E. D. Williams, A. L. Roytburd and R. Ramesh, "Role of substrate on the dielectric and piezoelectric behavior of epitaxial lead magnesium niobate-lead titanate relaxor thin films" *Appl. Phys. Lett.* **77** (2000) 438-440.

19. R. Herdier, M. Détalle, D. Jenkins, D. Remiens, D. Grébille and R. Bouregba, "The properties of epitaxial PMNT thin films grown on SrTiO$_3$ substrates" *J. Cryst. Growth*, **311** (2008) 123–127.

20. K.Y. Chan, W.S. Tsang, C.L. Mak and K.H. Wong, "Effects of composition of PbTiO3 on optical properties of (1-x)Pb(Mg$_{1/3}$Nb$_{2/3}$)O$_3$−xPbTiO$_3$ thin films", *Phys. Rev. B*, **69** (2004) 144111-1–144111-5.

21. K. L. Saenger, R. A. Roy, D. B. Beach, K. F. Etzold, "Pulsed Laser Deposition of High-Epsilon Dielectrics: PMN and PMN-PT", *Mater. Res. Soc. Symp. Proc.*, **285** (1993) 421-426.
22. P. Kumar, Sonia, R.K. Patel, C. Prakash, T. C. Goel, "Effect of substrates on phase formation in PMN-PT 68/32 thin films by sol–gel process", *Mater. Chem. and Phys.* **110** (2008) 7–10.
23. S. Jaydeep, S. Yadav, B. P. Malla, A. R. Kulkarni, and N. Venkatramani, "Growth and dielectric behavior of radio frequency magnetron-sputtered lead magnesium niobate thin films" *Appl. Phys. Lett.* **81** (2002) 3840-3842.
24. M. Ohring, "Material Science of Thin Films", Academic Press, London, 2$^{nd}$ Edition, pp 523.
25. Z. Zhao, V. Busaglia, M. Viviani, M. T. Busaglia, L. Mitoserin, A. Testino, M. Nygreen, M. Johnson and P. Nanni, "Grain-size effects on the ferroelectric behavior of dense nanocrystalline $BaTiO_3$ ceramics", *Phys. Rev. B*, **70** (2004) 024107-1–024107-8.
26. A. K. Tagantsev, M. Landivar, E. Colla, and N. Setter, "Identification of passive layer in ferroelectric thin films from their switching parameters", *J. Appl. Phys.*, **78** (1995) 2623-2630.
27. T. C. Goel, P. Kumar, A. R. James and C. Prakash, Processing and Dielectric Properties of Sol-Gel Derived PMN-PT(68:32) Thin Films, *J. Electroceram.*, **13** (2004) 503–507.
28. W. Z. Li, J. M. Xue, Z. H. Zhou, J. Wang, H. Zhu, J. M. Miao, $0.67Pb(Mg_{1/3}Nb_{2/3})O_3$–$0.33PbTiO_3$ thin films derived from RF magnetron sputtering, *Ceram. Intern.* **30** (2004) 1539–1542.
29. M. Detalle, G. Wang, D. Remiens, P. Ruterana, P. Roussel, B. Dkhil, Comparison of structural and electrical properties of PMN-PT films deposited on Si with different bottom electrodes, *J. Cryst. Growth*, **305** (2007) 137–143.
30. X. Y. Chen, K. H. Wong, C. L. Mak, J. M. Liu, X. B. Yin, M. Wang, Z. G. Liu, Growth of orientation-controlled $Pb(Mg_{1/3}Nb_{2/3})O_3$−$PbTiO_3$ thin films on Si(100) by using oriented MgO films as buffers, *Appl. Phys. A* **74** (2002) 1145-1149.
31. S.B. Ren, C.J. Lu, J.S. Liu, H.M. Shen, Y.N. Wang, Size-related ferroelectric-domain-structure transition in a polycrystalline $PbTiO_3$ thin film, *Phys. Rev. B* **54** (1996) 14337 14340.

32. S. Chattopadhyay, P. Ayyub, V.R. Palkar, M. Multani, Size-induced diffuse phase transition in the nanocrystalline ferroelectric PbTiO$_3$, *Phys. Rev. B* **52** (1995) 13177-13184.
33. G. Arlt, D. Henngings, G. de With, Dielectric properties of finegrained barium titanate ceramics, *J. Appl. Phys.* **58** (1985) 1619-1625.

# Table Captions

Table 1: Comparison of dielectric and ferroelectric properties of PMN-PT thin film on glass, Pt/SiO$_2$/Si and buffered Pt/SiO$_2$/Si substrates deposited by different methods

# Figure Captions

**Figure 1.** X-ray diffraction patterns of as-deposited and annealed PMN-PT films. For comparison X-ray diffraction pattern of PMN-PT target used for sputtering is also shown.

**Figure 2.** Cross-sectional FE-SEM image of PMN-PT film.

**Figure 3.** Surface FE-SEM images of PMN-PT films annealed at (a) 450 °C (b) 550 °C (c) 650 °C.

**Figure 4.** Frequency dependence of dielectric constant ($\varepsilon_r$) and dissipation factor (tan $\delta$) of PMN-PT films annealed at different temperatures.

**Figure 5.** P-E hysteresis loops of 550 °C and 650 °C annealed films recorded at different frequencies.

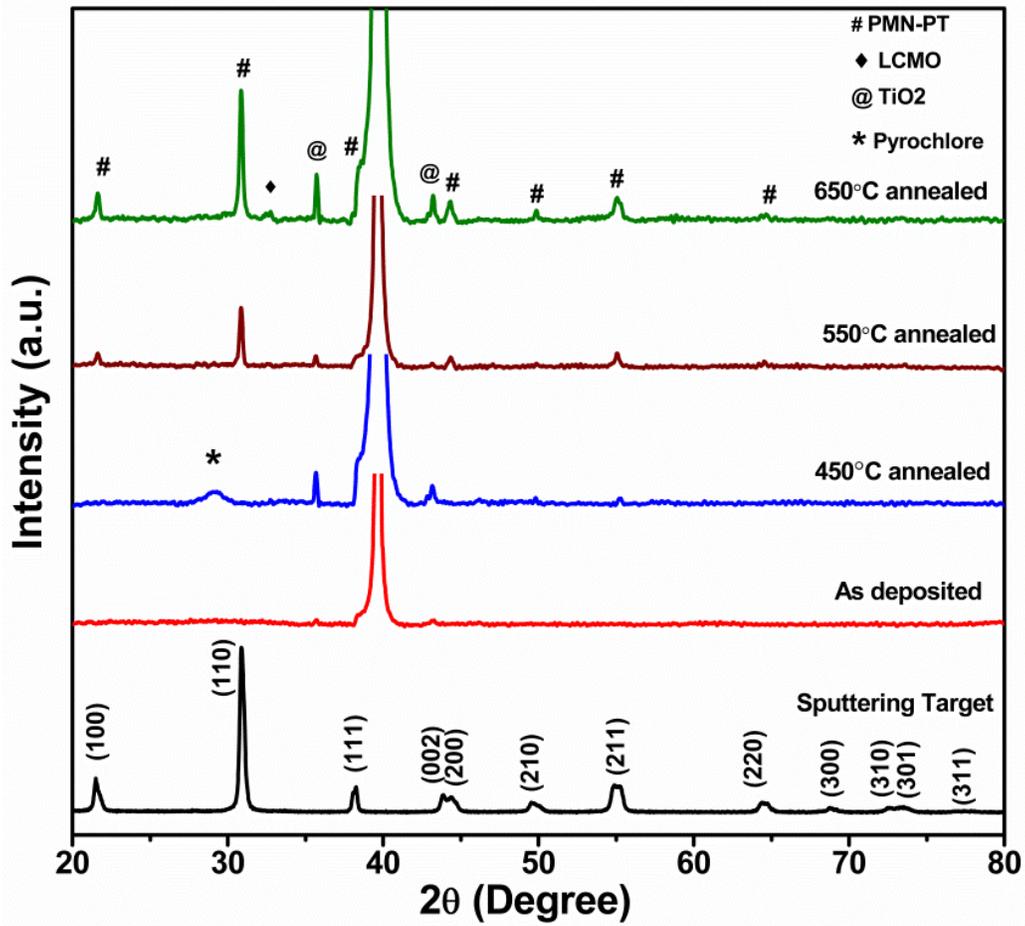

Figure 1.

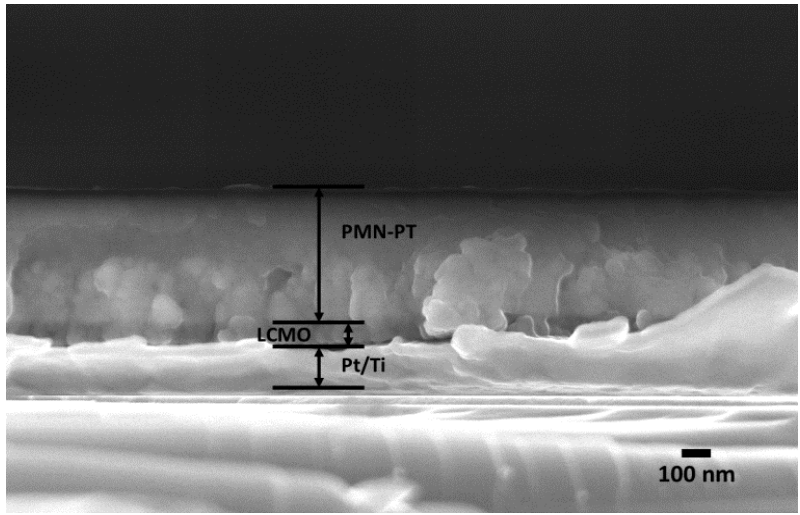

Figure 2.

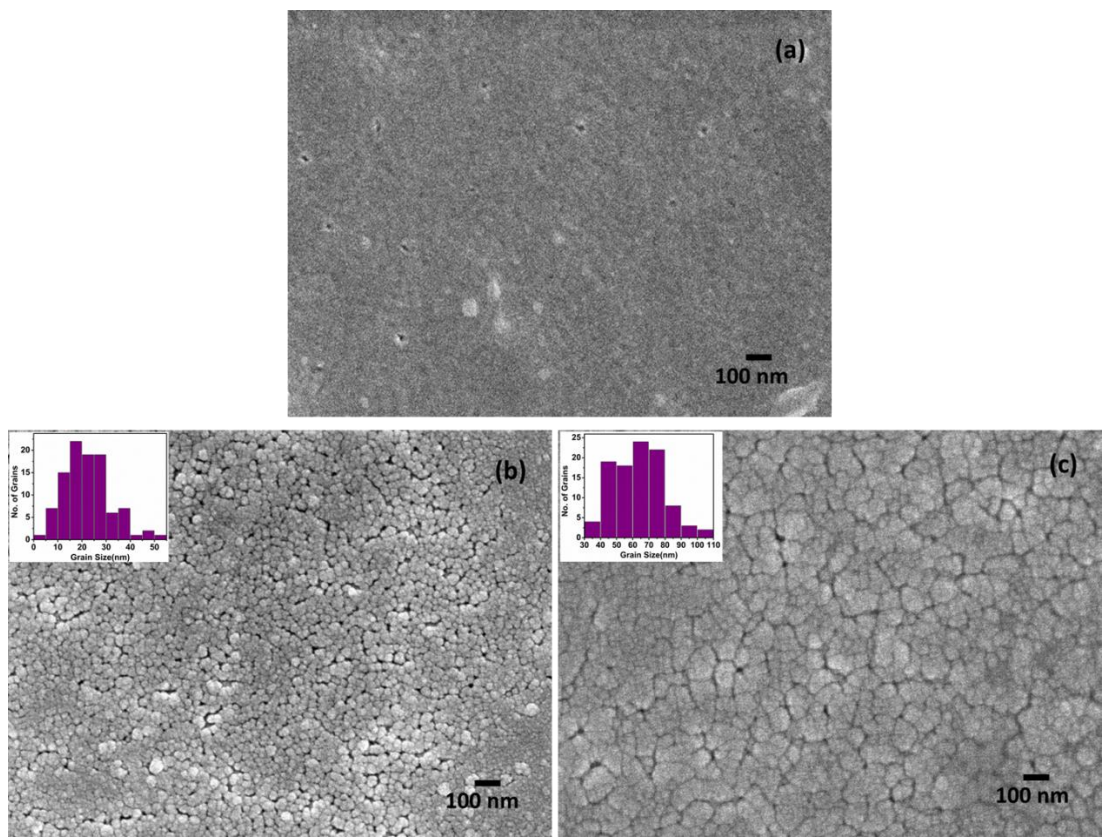

Figure 3.

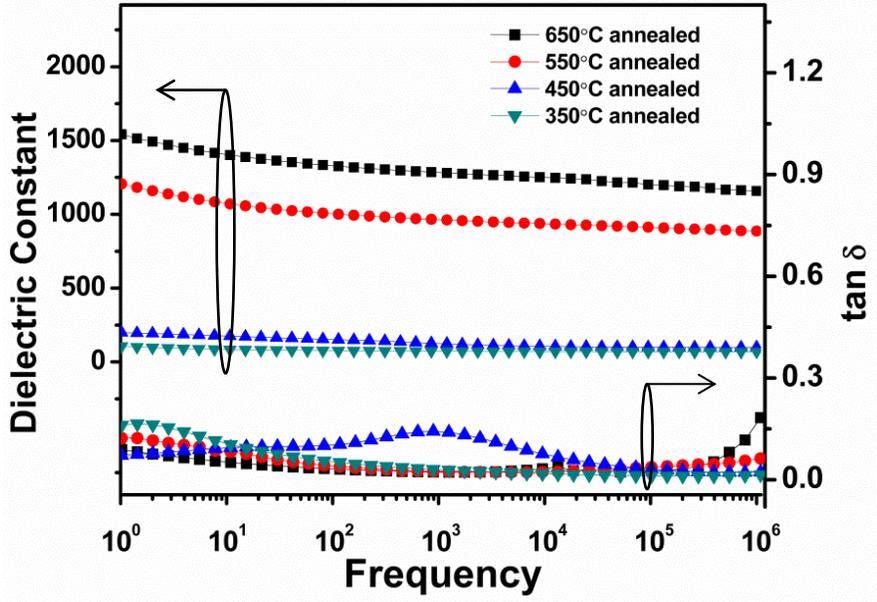

Figure 4.

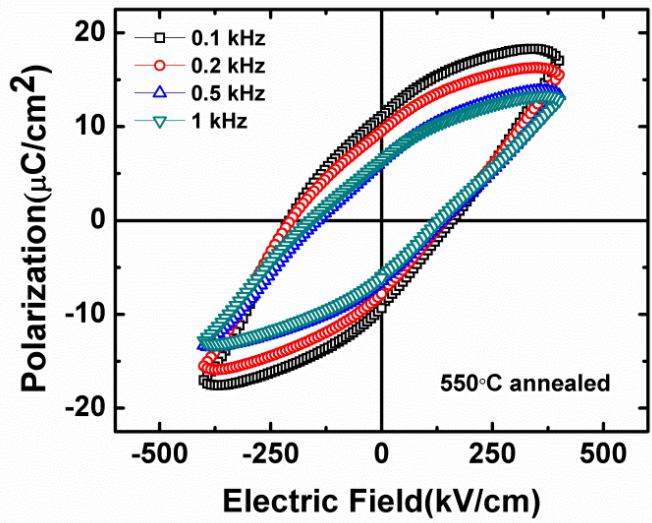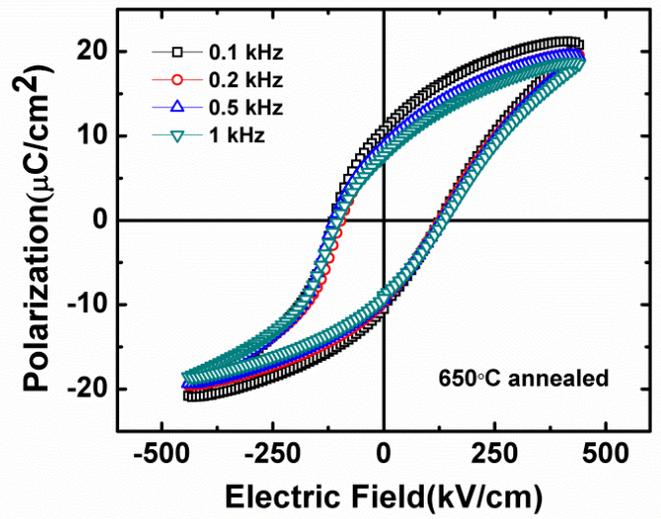

Figure 5.

Table 1.

| Composition | Substrate | Deposition method | Single Phase | $\varepsilon_r$ at 1kHz | $P_s$ ($\mu C/cm^2$) | Thickness (nm) | Processing temperature $T_s$ or $T_a$(°C) | Ref. |
|---|---|---|---|---|---|---|---|---|
| 65PMN-35PT | LCMO/Pt/TiO$_2$/Glass | RF sputtering | Yes | 1300<br>1000 | 22<br>18 | 500 | 650<br>550 | Present work |
| PMN | Au/Cr/Glass | RF Sputtering | Yes | 2052 | 24 | 800 | 500 | 23 |
| 68PMN-32PT | ITO/Glass,<br>Platinized Silicon | Sol-Gel | No<br>No | 950<br>835 | 2<br>2 | 800 | 600 | 27 |
| 67PMN-33PT | Pt/Ti/SiO$_2$/Si | RF Sputtering | Yes | 1090 | 46 | 500 | 550 | 28 |
| 70PMN-30PT | Pt/TiO$_x$/SiO$_2$/Si,<br>LNO/SiO$_2$/Si | RF Sputtering | No | 750<br>950 | 35<br>34 | 600 | 450 | 29 |
| 70PMN-30PT | Pt/Ti/SiO$_2$/Si | RF Sputtering | Yes | 2200 | - | 1440 | 640 | 7 |
| 70PMN-30PT | LNO/MgO/Si | PLD | Yes | 1350 | 34 | 600 | | 30 |
| 90PMN-10PT | PT/Pt/SiO$_2$/Si,<br>BT/Pt/SiO$_2$/Si,<br>BST/Pt/SiO$_2$/Si | PLD | No | 1043<br>1094<br>1274 | -<br>-<br>- | 500 | 530 | 9 |
| 65PMN-35PT | Pt/SiO$_2$/Si | PLD | No | 600-1200 | - | 600 | 650 | 21 |

LCMO: La$_{0.67}$Ca$_{0.33}$MnO$_3$, ITO: Indium tin oxide, LNO: LaNiO$_3$, PT:PbTiO$_3$, BT: BaTIO$_3$, BST: Ba$_{0.8}$Sr$_{0.2}$TiO$_3$, T$_s$: Substrate temperature, T$_a$: Annealing temperature